\documentclass{osa-article}
\journal{josab}

\articletype{Research Article}
\begin{document}

\title{Inhomogeneous nonlinearity meets $\mathcal{PT}$-symmetric Bragg structures: Route to ultra-low power steering and peculiar stable states}

\author{S. Sudhakar,\authormark{1} S. Vignesh Raja, \authormark{2} A. Govindarajan, \authormark{3,*} K. Batri, \authormark{4} and M. Lakshmanan
\authormark{3}}

\address{\authormark{1}Department of Electronics and Communication Engineering, Vel Tech High Tech Dr. Rangarajan Dr. Sakunthala
Engineering College, Chennai, 600 062, India\\
\authormark{2}Department of Physics, Pondicherry University, Puducherry, 605014, India\\
\authormark{3}Department of Nonlinear Dynamics, School of Physics, Bharathidhasan University, Tiruchirapalli, 620024, Tamilnadu,
India \\\authormark{4}Department of Electronics and Communication Engineering, PSNA College of Engineering and Technology, Dindigul, 624
622, Tamilnadu, India}

\email{\authormark{*}Corresponding author:govin.nld@gmail.com} 

\begin{abstract}
 In the context of $\mathcal{PT}$-symmetric fiber Bragg gratings, tailoring the nonlinear profile along the propagation coordinate serves to be a new direction for realizing low-power all-optical switches. The scheme is fruitful only when the nonlinearity profile will be either linearly decreasing or increasing form. If the rate of variation of the nonlinearity profile is high, the critical intensities fall below the input power of value 0.01 in the unbroken regime provided that the light launching direction is right. Nowadays, every new theoretical inception into the PTFBG has started making sense of switching in the broken $\mathcal{PT}$-symmetric regime which was once believed to be the instability regime. When the inhomogeneous nonlinearity acts together with the broken $\mathcal{PT}$-symmetry and right light incidence, it leads to two peculiar settings. First, the switch-up intensities are ultra-low. Second, the switch-down action takes place at zero critical intensities. Such OB curves are unprecedented in the context of conventional gratings and found only in plasmonic devices and anti-directional couplers.  Even though the nonlinearity is inhomogeneous, the ramp-like first stable states persist in the broken $\mathcal{PT}$-symmetric regime giving an additional indication that the broken PTFBG is closely associated with the plasmonic structures.   In the existing PTFBG systems, the switching intensities are relatively higher in the broken regime. However, the proposed system records the lowest switching intensities in the broken regime. The reported intensities ($< 0.005$) are also the lowest ever-switching intensities recorded in the perspective of PTFBGs to date.  
\end{abstract}

%%%%%%%%%%%%%%%%%%%%%%%%%%  body  %%%%%%%%%%%%%%%%%%%%%%%%%%
\section{Introduction}
	Fabricating fiber Bragg grating (FBG) demands exposing and masking neighboring regions of fiber core to extreme UV radiation. The refractive index (RI) of the exposed areas gets altered by the exposure to UV. The masked areas do not feature any such RI variations. In this fashion, the RI of the adjacent regions mismatches with one another. The change in the RI of the regions exposed to UV is very marginal ($\Delta n$ = $n-n_0$) compared to  the RI of the masked regions of the core ($n_0$). Even though the RI variation is marginal, it is sufficient to make the device reflective to a band of wavelengths known as the \emph{stopband} of the grating. Building applications from simple filters to advanced all-optical signal processing devices with FBGs require suitable customization of this stopband \cite{giles1997lightwave, othonos1997fiber,hill1997fiber,erdogan1997fiber}. Introducing nonlinearity is one of the schemes to alter the stopband characteristics of FBG, and this has emerged as a separate research branch in the context of modern fiber optics \cite{agrawal2001applications,de1990switching,winful1979theory,herbert1993optical}.
	
	In the context of nonlinear FBGs, several theoretical approaches have been suggested  by researchers towards the development of low-power all-optical switches (AOS) and signal processing devices.  From a material perspective, FBGs inscribed on highly nonlinear fibers offer steering at low critical intensities (switch-up and switch-down) \cite{karimi2012all, harbold2002highly,chen2006measurement,yosia2007double,yousefi2015all}. These intensities get reduced with the optimization in the signal parameters. For instance, if the wavelength of the continuous wave (CW) pump source is detuned far away from the Bragg and operating wavelength of the FBG, the intensities required to steer between the switching states  are effectively reduced \cite{zang2012,broderick1998bistable,broderick1998nonlinear,ping2005bistability}. Realization of low-power AOS in the presence of Gaussian and square-shaped pulses demands optimization of the width of the launched pulses alongside the length and coupling coefficient of the FBG \cite{ping2005nonlinear,lee2003nonlinear,de2006ultrashort}. Inhomogeneities in the form of chirping introduced into an FBG tailors the switching intensities to the required levels besides broadening the spectral span and increasing the spectral uniformity \cite{kim2001effect,maywar1998effect,maywar1998low}.  From the design perspective,  introducing a $\pi$-phase-shift region in the middle of the FBG is reckoned to be one of the effective methodologies proposed so far as the design of low-power AOS is concerned \cite{radic1994optical,radic1995theory,suryanto2009numerical}. 	The directions to control the switch intensities in a nonlinear FBG are not limited to these techniques alone.

 The notion of inhomogeneous nonlinearity is emerging as a distinct route for making all-optical devices such as switches, logic gates, and other signal processing devices \cite{pinto2017bistability,nobrega1998optimum,nobrega2000multistable,sales2015mach,coelho2013switching,sobrinho2002acousto,fraga2006all,liang2005nonlinear,filho2020nonlinearity}.  As the name implies, the term inhomogeneous nonlinearity refers to the variation in the nonlinearity ($\gamma$) as a function of propagation coordinate $z$. In other words, the value of the nonlinear coefficient is not constant, but it varies as the light traverses through the device [$\gamma(z)$]. From a theoretical standpoint, scientific works which deal with the switching dynamics of anti-directional couplers (ADCs) \cite{coelho2013switching}, and directional couplers (DCs) \cite{nobrega1998optimum,nobrega2000multistable,filho2020nonlinearity} with inhomogeneous nonlinear profiles are many. There seem to exist a few works dealing with the switching in a grating with inhomogeneous nonlinearity. Compared to the conventional FBG devices with the homogeneous nonlinearities \cite{pinto2017bistability}, gratings with varying nonlinear profiles switch at higher or lower critical intensities depending on the profile used and the value of the inhomogeneous nonlinearity parameter ($\gamma(\zeta)$).    Different kinds of nonlinear profiles exist in the form of linear,  Gaussian, exponential, logarithmic, and so forth \cite{coelho2013switching,nobrega1998optimum,nobrega2000multistable,filho2020nonlinearity,liang2005nonlinear}.  Judicious selection of the nonlinearity profile and manipulation of the  inhomogeneous nonlinearity parameter are necessary for acquiring low-power AOS.
 
  The realization of inhomogeneous nonlinearity may require variation in the doping level as a function of the propagation coordinate. This is because the nonlinear parameter ($\gamma$) is a function of the nonlinear coefficient of the material. The same material with a wide range of nonlinear coefficients can be manufactured by varying the doping levels. It is possible to produce a fiber with an inhomogeneous nonlinear profile by suitably controlling the thickness, the density of the dopant, or dopant diffusion time \cite{coelho2013switching,nobrega1998optimum,nobrega2000multistable,filho2020nonlinearity,liang2005nonlinear}. Controlling of these parameters independently produces inhomogeneity in the linear RI profile ($n_0$). Theoretical physicists attempted to quantify the correlation between inhomogeneously varying linear RI and nonlinear RI ($n_2$) via bond-orbital theory. The mathematical relationship between linear and nonlinear RI in terms of the Sellmeier gap ($\mathcal{E}_s$), bond length ($d_l$) and cation valence ($\mathcal{Z}$) reads as
  \begin{gather}
  	n_2=\cfrac{\mathcal{Z}\left(n_0^2-1\right)d_l^2}{n_0 \mathcal{E}_s^2 }10^{-16}\left(\cfrac{cm^2}{W}\right)
  	\label{Eq:0}
  \end{gather}
	From Eq.(\ref{Eq:0}), it is clear that the material should have high/low linear RI, large (small) bond length, or small (large) Sellmeier gap, to exhibit large (small) value of nonlinear coefficient ($n_2$). Thus, it is possible to fabricate fiber that support inhomogeneous nonlinearity via variable metal doping \cite{coelho2013switching,nobrega1998optimum,nobrega2000multistable,filho2020nonlinearity,liang2005nonlinear}.

	 The past decade has marked substantial developments in the field of $\mathcal{PT}$-symmetry ever since the translation of the original concepts from quantum to classical systems \cite{ruter2010observation,govindarajan2018tailoring,feng2017non,longhi2018parity,kottos2010optical}. It is well known that any conventional photonic device including the FBG can be turned into a $\mathcal{PT}$-symmetric device by supplying a balanced amount of gain and loss \cite{regensburger2012parity,el2007theory,el2018non,ozdemir2019parity,article}. Even before the formulation of $\mathcal{PT}$-symmetric concepts in optics, the approach to place gain regions next to the lossy regions was conceptualized in the context of FBGs by Poladian \cite{poladian1996resonance} and Kulishov \emph{et al} \cite{kulishov2005nonreciprocal}. Under the periodic modulations of the real part of RI and gain-loss profile, FBGs have been examined intensively in the linear regime \cite{lin2011unidirectional,raja2020tailoring,raja2020phase,raja2021n}. Phang \emph{et al} discovered that PTFBGs are ideal candidates for developing ultrafast AOS and signal processing devices \cite{phang2013ultrafast,phang2014impact,phang2015versatile}.  The concept of nonreciprocal switching in nonlinear FBGs under the  reversal of light launching direction was discovered by Komissarova \emph{et al} \cite{komissarova2019pt,shestakov2021peculiarities}, and it has been identified as the perfect solution for realizing low-power AOS by Raja \emph{et al} \cite{raja2019multifaceted,PhysRevA.100.053806}. In the process, switching dynamics of both homogeneous \cite{1555-6611-25-1-015102} and inhomogeneous PTFBGs were thoroughly examined in different $\mathcal{PT}$-symmetric regimes under the variation in the system and signal parameters. The inferences made from these works strongly emphasize that the detuning parameter should not be taken lightly as it plays a vital role in reducing the switching intensities \cite{raja2019multifaceted}. Apart from the low-power switching, the inhomogeneous PTFBG presents an OB curve with a high degree of spectral uniformity and tunable spectral span. The same authors also  corroborated that operating in the broken $\mathcal{PT}$-symmetric regime leads to the visualization of ramp-like stable states in the input-output characteristics. Surprisingly, these unconventional stable states appear for a wide range of system parameters \cite{PhysRevA.100.053806}. The quest to realize low-power AOS motivated us to study the switching dynamics of PTFBG with modulation of Kerr nonlinearity. In the same numerical experiment, the lowest switching intensities in the context of PTFBGs have occurred in the unbroken $\mathcal{PT}$-symmetric regime thanks to the idea of reversal of the direction of light incidence \cite{sudhakar2022low}. 
	 
	 In the search for alternative routes to design a low-power AOS, we investigate the nonreciprocal switching dynamics of a PTFBG with inhomogeneous nonlinearity. The inhomogeneous nonlinearity [$\gamma(z)$] may increase or decrease along the propagation length. But in this article, we will restrict ourselves to the study of $\mathcal{PT}$-symmetric FBG (PTFBG) with the linearly decreasing nonlinearity profile and its mathematical description is given in Sec. \ref{model}. In the Section that follows, the OB exhibited by a PTFBG with the linearly decreasing nonlinearity profile is presented in different operating regimes, namely unbroken and broken, for both the light incident directions. In the subsequent section, the major highlights of our numerical investigation are summarized.
	 
\section{Theoretical framework}
\label{model}
A typical nonlinear FBG will feature only third-order (cubic) nonlinearity. The nonlinearity parameter ($\gamma$) is a function of the nonlinear coefficient $n_2$. Inscribing a grating requires modulation of the real part of RI ($n_{1I}$) of a fiber. Such a modulation dictates the coupling strength ($\kappa$) between the counter propagating waves and the reflectivity. It is assumed that the RI of the bare fiber without grating is $n_0$. To establish $\mathcal{PT}$-symmetry, gain-loss modulation parameter ($g$) should be included into the FBG's RI profile via the imaginary part of modulation of RI ($n_{1I}$). The effective RI of the proposed system is thus written as
\begin{gather}
	n(z)=n_0+n_{1R} cos(2 \pi z/\Lambda)+in_{1I} sin(2 \pi z/\Lambda)+n_2(z)|E|^2.
	\label{Eq:1}
\end{gather}
In Eq. (\ref{Eq:1}), $\Lambda$ stands for the grating period and $|E|^2$ signifies the strength of the incoming optical field. The total electric field propagating inside the FBG is the sum of forward ($E_f$) and backward propagating fields $E_b$ and it reads as
\begin{gather}
	E(z)=E_f \exp(ikz)+E_b \exp(-ikz).
		\label{Eq:2}
	\end{gather}
	In Eq. (\ref{Eq:2}), $k$ stands for the wave vector.  It should be noted that Eq. (\ref{Eq:1}) obeys the $\mathcal{PT}$-symmetric condition $n(z)=n^*(-z)$. The Helmholtz equation that governs the propagation of electric field in a optical medium is given by,
	\begin{gather}
		\cfrac{d^2E}{dz^2}+k^2\left(\cfrac{n^2(z)}{n_0^2}\right)=0.
		\label{Eq:3}
	\end{gather} 
Substituting Eqs. (\ref{Eq:1}) and (\ref{Eq:2}) in (\ref{Eq:3}) and by using transformation $E_{f, b}$ = $A_{f, b}$ $\exp(\mp\delta_0 z)$, we arrive at the coupled-mode equations (CMEs) that govern the propagation of light in a PTFBG with the inhomogeneous cubic-nonlinearity  as

	\begin{gather}
		+i\frac{d A_{f}}{dz}+\delta_{0} A_{f}+\left(k_{0}+g_{0}\right)A_{b}+\gamma_{0}(z)\left(|A_{f}|^{2}+2|A_{b}|^{2}\right)A_{f}=0,
		\label{Eq:4}
	\end{gather}
	\begin{gather}
		-i\frac{d A_{b}}{d z}+\delta_{0}A_{b}+\left(k_{0}-g_{0}\right)A_{b}+\gamma_{0}(z)\left(|A_{b}|^{2}+2|A_{f}|^{2}\right)A_{b}=0.
		\label{Eq:5}
		\end{gather}
It should be noted that the higher order derivatives of electric fields are neglected in the  CMEs using a slowly varying envelope approximation. From the fundamentals of FBG, the relation between various system parameters in CMEs (\ref{Eq:4}) and (\ref{Eq:5}) with the parameters in Eq. (\ref{Eq:1}) and operating wavelength ($\lambda$) are defined as
\begin{gather}
\nonumber	\delta_0=2 \pi n_0\left(\cfrac{1}{\lambda}-\cfrac{1}{\Lambda_b}\right), \quad \lambda_b=2 n_0 \Lambda, \quad \kappa_0 = \pi n_{1R}/\lambda, \\ g_0 = \pi n_{1I}/\lambda, \quad \gamma_0=2 \pi n_2/\lambda
\label{Eq:6}
\end{gather}
The parameter $\delta_0$ determines the amount of detuning of the operating wavelength from the Bragg wavelength ($\lambda_b$). A normalization routine is required at this juncture to transform the variables from physical to normalized units, which as follows
\begin{gather}
	{\zeta}=\cfrac{z}{z_{0}},\quad  \psi_f=\cfrac{A_{f}}{\sqrt{P(0)}},\quad \psi_b=\cfrac{A_{b}}{\sqrt{P(0)}},
	\label{Eq:7}
\end{gather}
where $P(0)=|\psi_f(0)|^2$ is the intensity of the input laser.  Also, $z_{0}$ is the length scale over which the physical units are normalized. The normalized propagation coordinate, forward, backward and output field intensities are given by $\zeta$, $\psi_f$, $\psi_b$, and $P_1(L)=|\psi_f|^2$, respectively. The effect of normalization impacts on the system parameters as well and the normalized system parameters as
\begin{gather}
	\delta={\delta_{0}z_{0}},\quad \kappa=\kappa_{0}z_{0},\quad
	g={g_{0}} z_{0}, \quad
	\gamma(z)={\gamma_{0}(z)} P(0) z_{0}.
	\label{Eq:8}
\end{gather}
Substituting these normalized variables in Eqs. (\ref{Eq:8}) and (\ref{Eq:7}) in CMEs (\ref{Eq:4}) and (\ref{Eq:5}),  we obtain
\begin{gather}
	+i\frac{d \psi_f}{d\zeta}+{\delta}\psi_f+\left({k}+{g}\right)\psi_b+{\gamma(\zeta)}\left(|\psi_f|^{2}+2|\psi_b|^{2}\right)\psi_f=0,\label{Eq:norm1}\\
	-i\frac{d \psi_b}{d\zeta}+{\delta}\psi_b+\left({k}-{g}\right)\psi_f+{\gamma(\zeta)}\left(|\psi_b|^{2}+2|\psi_f|^{2}\right)\psi_b=0. \label{Eq:norm2}
\end{gather}
The inhomogeneous nonlinear profile is assumed to be linearly changing at a rate defined by
\begin{gather}
	\gamma(\zeta)=\begin{cases}1+\cfrac{(1-\sigma)\zeta}{ L}, & \text{linearly decreasing nonlinear coefficient} \\\\1+\cfrac{(\sigma-1)\zeta}{L}, & \text{for: linearly increasing nonlineary profile}\end{cases}
	\label{Eq:9}
	\end{gather}
 where $L$ is the normalized length of the PTFBG and $\sigma$ signifies the rate at which the nonlinearity is varying along the propagation coordinate ($\zeta$). 
 
 The system of equations in (\ref{Eq:norm1}) and (\ref{Eq:norm2}) are solved by implicit Runge-Kutta fourth order method. The boundary conditions to solve the CMEs are assigned as
 \begin{gather}
 	\psi_f(0)=\psi_f^0 \quad \text{and} \quad \psi_b(L)=0.
 \end{gather}
\section{Numerical Results}
\label{result}
The theory of optical bistability (OB) and optical multistability (OM) proposed by Winful \emph{et al.} lays the ground for the investigation of steering dynamics in nonlinear FBGs and hence a short recap of the physical phenomenon is required \cite{winful1979theory,herbert1993optical}. The nonlinear fiber optic devices that feature opposite directionality of Poynting vectors (representing counter-propagating waves) will exhibit two (OB) or more stable states (OM) in its transmission characteristics for a given input power. The propagating waves must posses opposite phase velocity and energy flows for the OB phenomenon to occur. This is the physical reason why OB is uncommon in the context of nonlinear DCs and universal in the perspective of nonlinear ADCs \cite{govindarajan2020tunable,govindarajan2019nonlinear}. Nonlinear FBGs exhibit OB due to the presence of feedback offered by the short-grating \cite{winful1979theory} without which OB cannot occur as in the case of long-period gratings. As the value of the input intensity ($P_0$) is tuned, the output intensity [$P_1(L)$] increases gradually till a critical intensity value which we designate as critical switch-up intensity ($P_{cr}^{\uparrow}$). The range of output intensities [$P_1(L)$] corresponding to input intensities in the span $0\leq P_0 < P_{cr}^{\uparrow}$ is generally referred as the first stable branch of the OB curve.  Any marginal increase in the input intensity beyond $P_{cr}^{\uparrow}$ leads to a sudden leap in the output intensity marking the onset of switch-up phenomenon. The output intensity of the system admits linearly increasing behavior with a further increase in the value of input intensity and is designated as the second stable branch. It is often used in practice that the input intensity must be tuned in reverse to recover the complete S-shaped hysteresis curve. On doing so, the system does not go back to the first stable branch at $P_0=P_{cr}^{\uparrow}$. Instead, the switch-down phenomenon happens at a different critical intensity level which is different from $P_{cr}^{\uparrow}$ and hence it is designated as critical switch-down intensity $P_{cr}^{\downarrow}$. The path from $P_{cr}^{\downarrow}\leq P_0 <P_{cr}^{\uparrow}$ also forms the part of the second stable branch. The vertical separation (in terms of the output intensity) between these two stable branches is often indicated by an unstable branch which is visible in numerical experiments alone. 

When it comes to the selection of device parameters, care must be taken that the feedback of the system is sufficient for the OB and OM to occur. The feedback parameter ($\kappa L$) must be judiciously adjusted in numerical experiments for the following reasons: High value of $\kappa L$ results in high values of critical intensities and larger hysteresis which is not the optimum requirement for ideal AOS. Suppose, if the value of $\kappa L$ is too low, the insufficient feedback will hinder the formation of OB curve. Physically realizable FBGs feature length in the range of 1mm to 10cm and coupling coefficient in the range of 1 to 10 $cm^{-1}$, respectively. Therefore, numerical scientists can work with any normalized value of $\kappa L$ from 1 to 10. The value of length and coupling coefficient are fixed as $L=2$ and $\kappa=3$, respectively, throughout this article as they lead to the formation of stable OB and OM curves for a wide range of operating conditions.
\subsection{OB in the unbroken $\mathcal{PT}$-symmetric regime: Left incidence}
\begin{figure}[t]
	\centering	\includegraphics[width=0.4\linewidth]{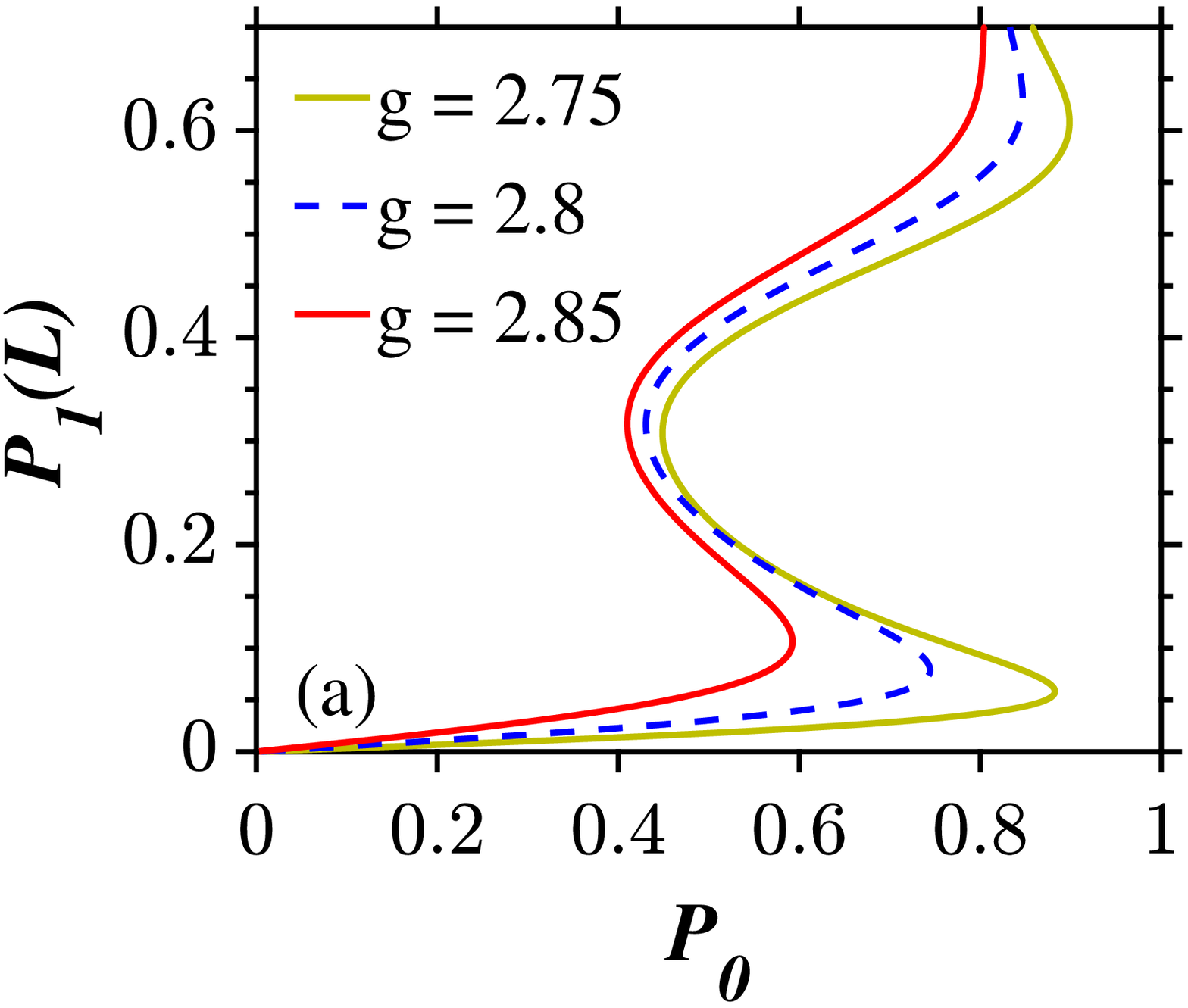}\centering	\includegraphics[width=0.4\linewidth]{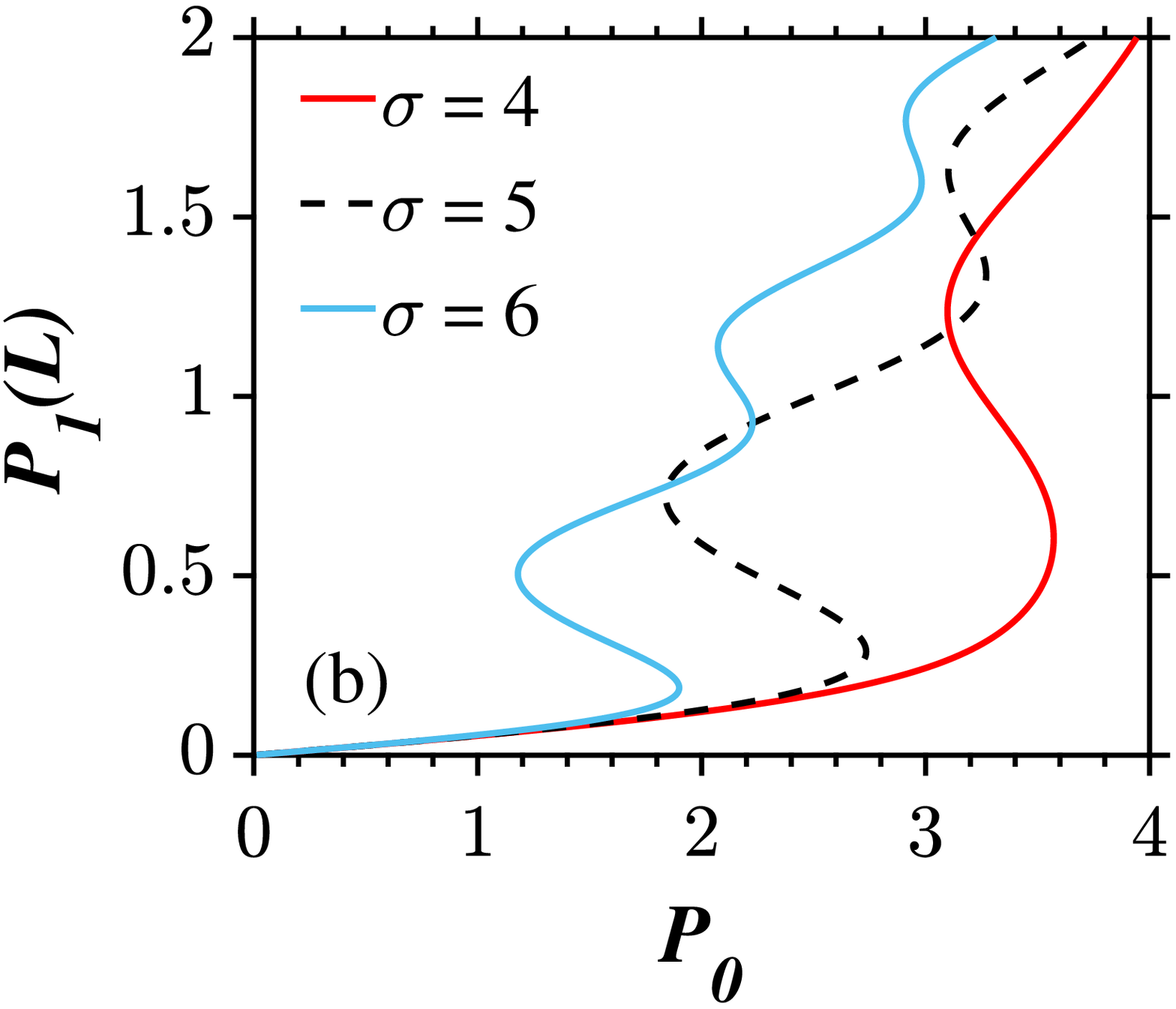}\\\includegraphics[width=0.4\linewidth]{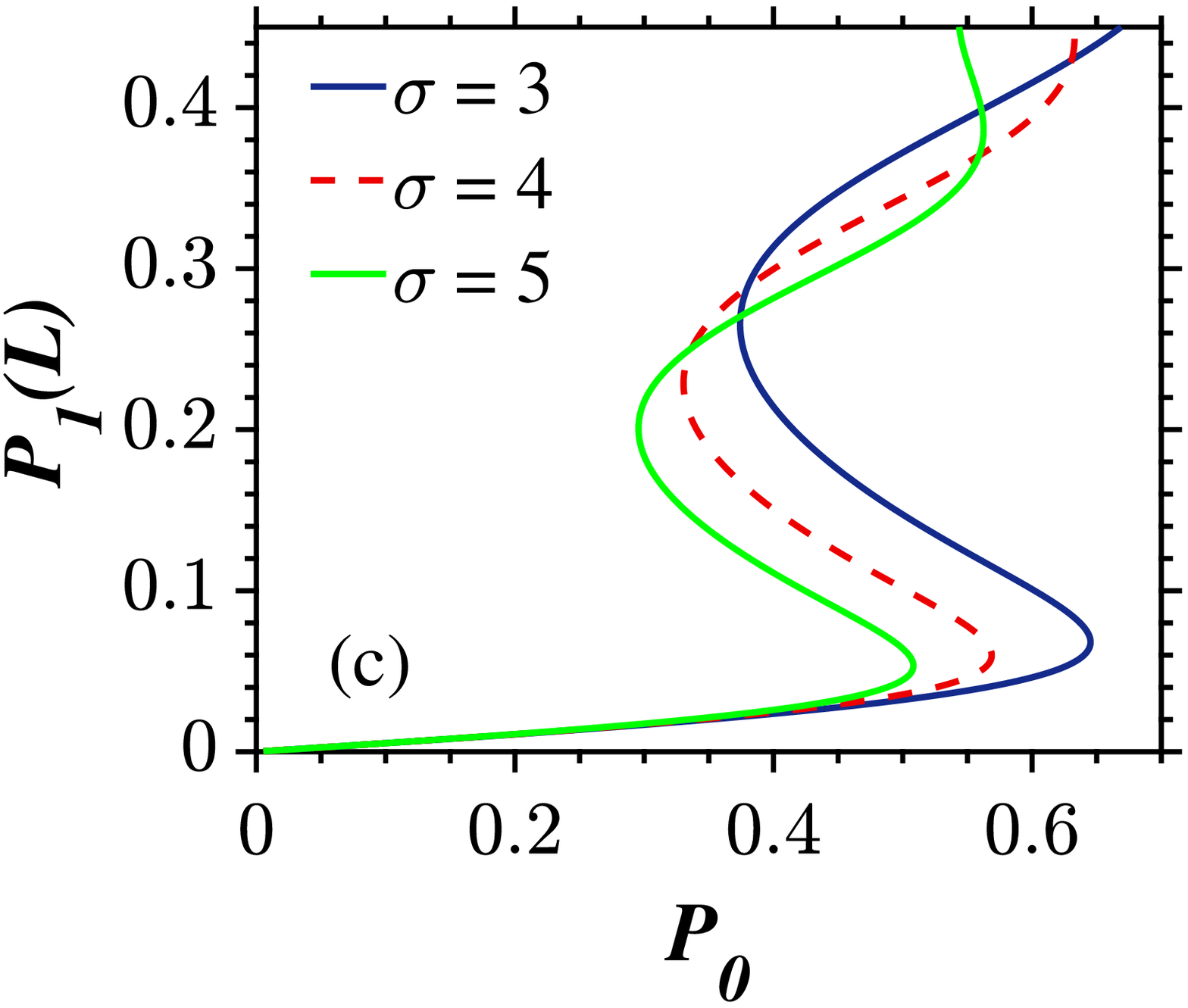}\includegraphics[width=0.4\linewidth]{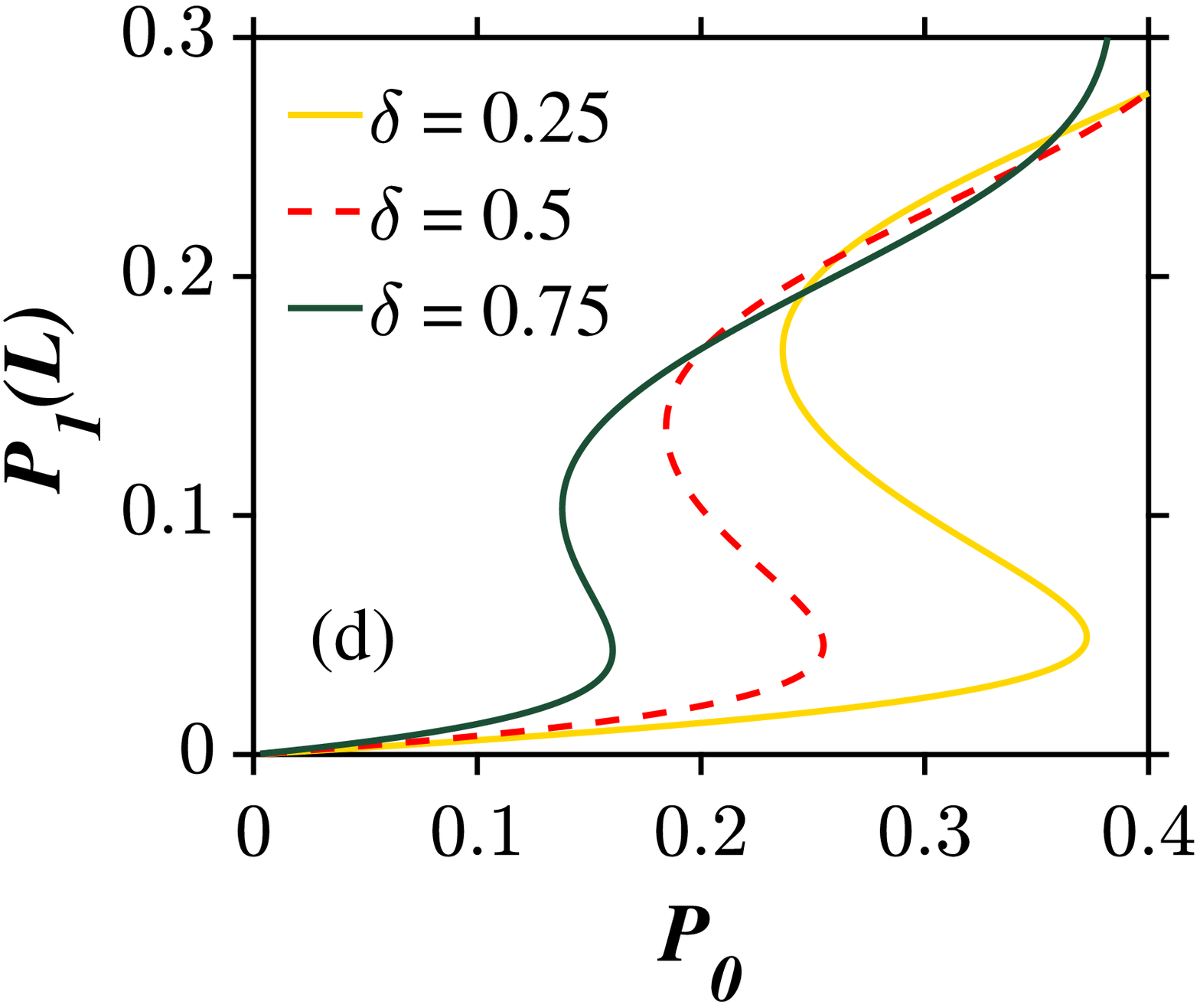}
	\caption{OB in unbroken PTFBGs ($g=2.8$) with an inhomogeneous nonlinearity. The light launching direction is left. The nonlinearity decreases linearly along $z$ in (a), (c), (d) and increases linearly in (b). (a)--(c) are obtained at the synchronous wavelength ($\delta=0$). Role of (a) gain and loss ($g$), (b),(c) inhomogeneous nonlinearity parameter ($\sigma$) and (d) positive detuning parameter ($\delta$) on the switching dynamics. In (a) and (d) $\sigma = 2$ and 5, respectively. }
	\label{fig1}
\end{figure}
The input-output characteristics of a PTFBG with the inhomogeneous nonlinearity operating in the unbroken regime reveals that any increase in the value of gain and loss parameter ($g$) leads to a decrease in the critical intensities ($P_{cr}^{\uparrow}$ and $P_{cr}^{\downarrow}$)  required for switching. Reduction in these critical intensities means that the width of the hysteresis ($\Delta_{\uparrow\downarrow}$) curve tends to narrow-down with higher values of gain and loss. It should be noted that the gain and loss parameter ($g$) serves as a useful parameter to reduce the switching intensities only when the value is not too far away from the unitary transmission point (2.6$\leq g\leq$2.9) as shown in Fig. \ref{fig1}(a). In Fig. \ref{fig1}(b), we have the results pertaining to the switching dynamics of unbroken PTFBG configuration where the inhomogeneous nonlinearity is subjected to variation in an increasing fashion in accordance with Eq. (\ref{Eq:9}). We notice that the critical intensities show a decreasing behavior against the variation in the inhomogeneous nonlinearity parameter ($\sigma$). However, the width of the hysteresis curve and switching intensities are higher in that case. For these reasons, one can argue that the linearly increasing inhomogeneous nonlinearity is not an ideal scenario for actualizing low-power AOS. The natural question arises here is that whether the linearly decreasing inhomogeneous nonlinearity profile will result in a different switching dynamics in the intended direction. To answer the query, the parameter $\sigma$ is numerically varied in Fig. \ref{fig1}(c) from 3 to 5. The OB curve corresponding to $\sigma = 5$ features lowest switching intensities as well as narrowest widths. This confirms the fact that the idea of decreasing  the nonlinearity inhomogeneously serves as an alternative route to realize low-power AOS. Operating the PTFBG away from the synchronous or Bragg wavelength ($\delta = 0$) assists in altering the shape and the critical switching intensities. Even though the PTFBG can be operated in both the longer and shorter wavelength sides of the Bragg wavelengths, the choice of the detuning regime should be made based on the type of nonlinearity. Recent works on nonlinear PTFBGs suggest that the device supports low power OB states in the negative detuning regime provided that the nonlinearity is a self-defocusing one \cite{raja2019multifaceted, PhysRevA.100.053806}. In the present work, the cubic nonlinearity is assumed to offer self-focusing effect and hence operating in the negative detuning regime ($\delta<0$) may not favor the formation of low power OB states. On the other hand, an increase in the value of detuning parameter in the positive detuning regime ($\delta>0$) leads to a decrease in the critical switching intensities as shown in Fig. \ref{fig1}(d). It should be remembered that the detuning parameter cannot be too large ($\delta>0.75$) because it leads to the loss of bistable states. 

\subsection{OB in the broken $\mathcal{PT}$-symmetric  regime: Left incidence} 
\begin{figure}
	\centering	\includegraphics[width=0.5\linewidth]{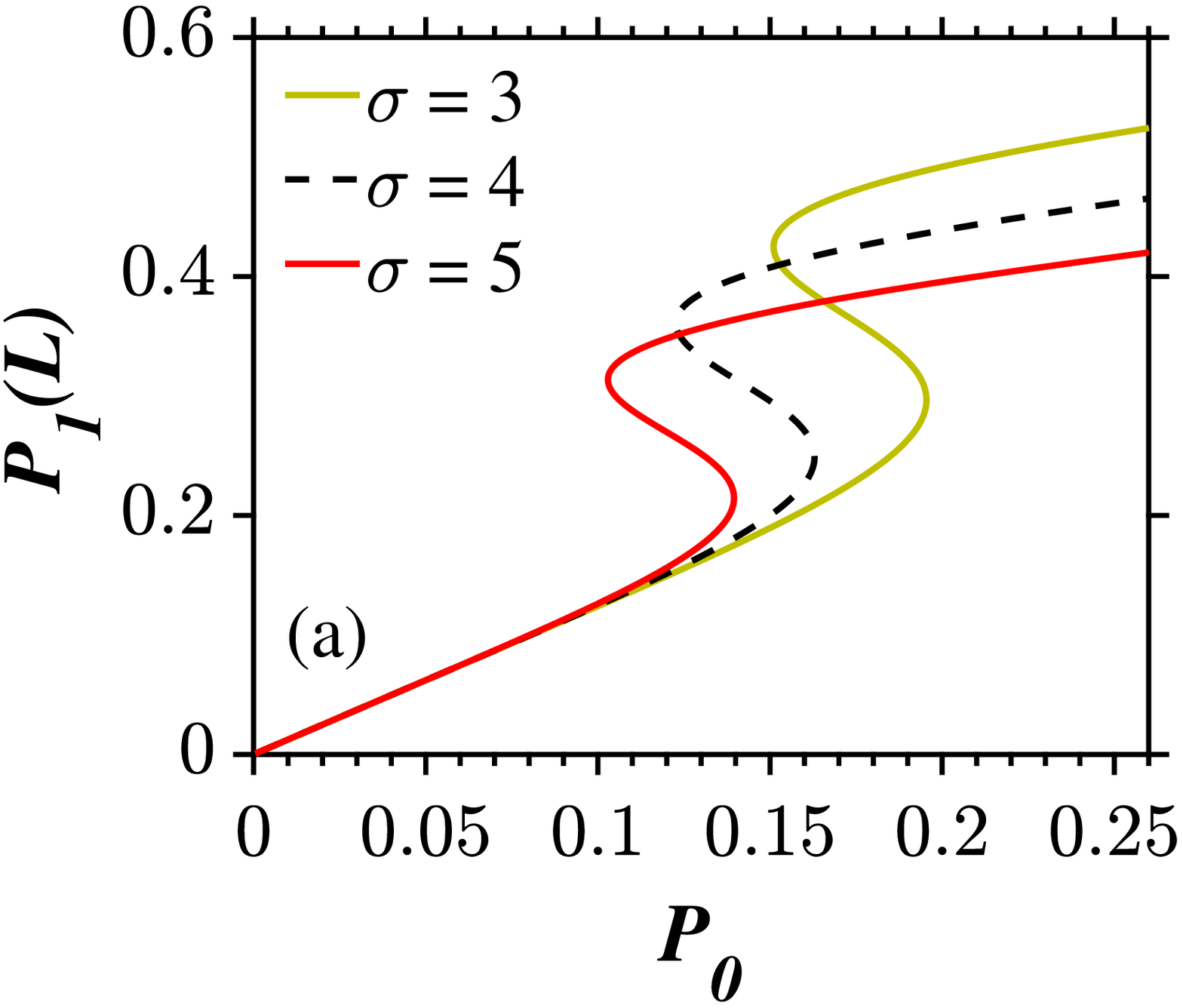}\centering	\includegraphics[width=0.5\linewidth]{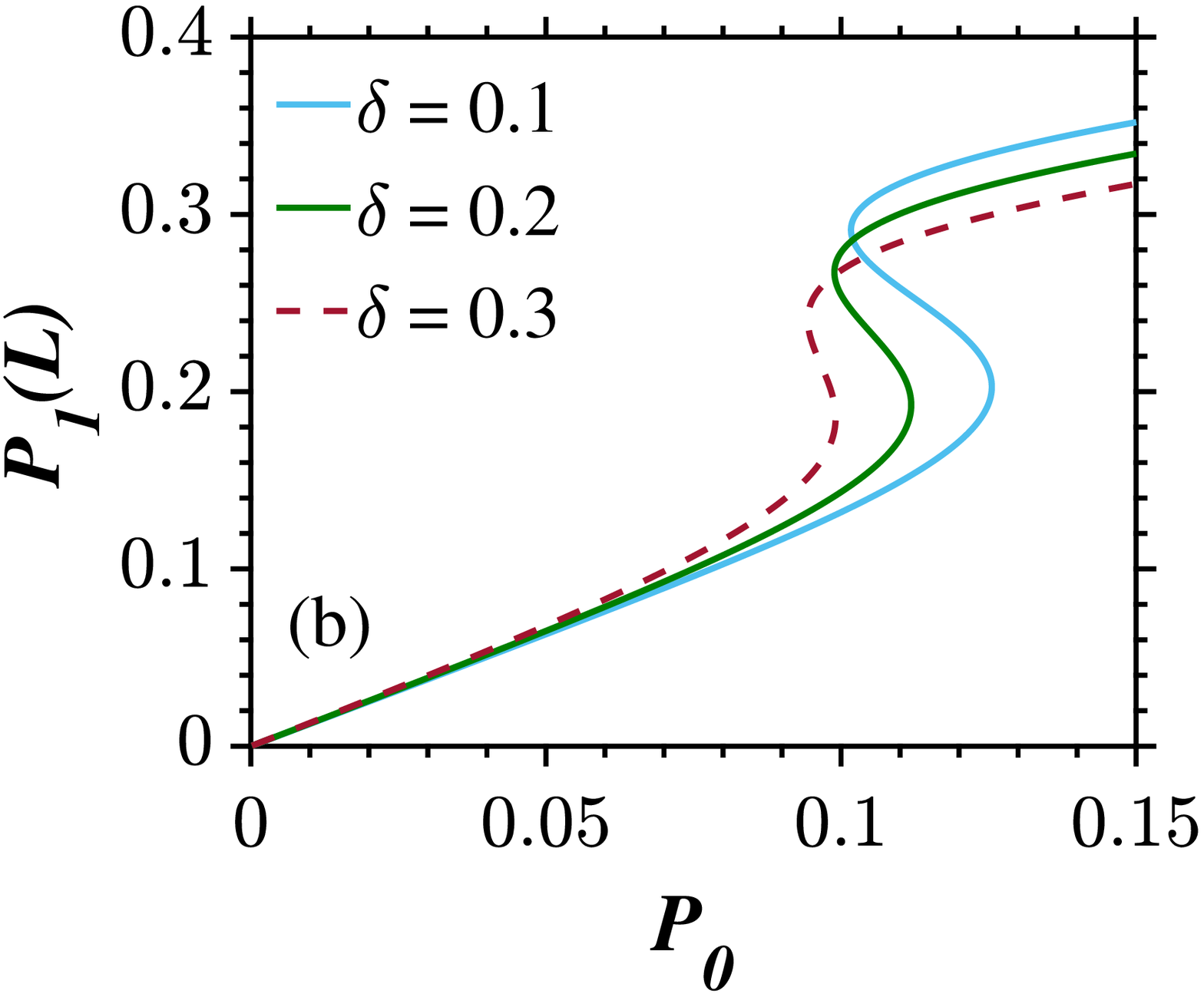}
	\caption{OB in  broken PTFBGs ($g=3.5$) with linearly increasing inhomogeneous nonlinearity under left light incidence direction. The OB curves features ramp-like first stable states. Role of (a) inhomogeneous nonlinearity parameter ($\sigma$) at $\delta = 0$ and (b) positive detuning parameter ($\delta$) at $\sigma = 5$ on altering the span of ramp-like states.}
	\label{fig2}
\end{figure}
Recent works on nonlinear PTFBGs suggest that operating the system in the broken $\mathcal{PT}$-symmetric regime yields bistable curves with unconventional ramp-like first stable branch \cite{raja2019multifaceted,PhysRevA.100.053806,sudhakar2022low}. The occurrence of ramp-like stable states is an uncommon feature in the conventional nonlinear FBG. But these kinds of curves are very common in nonlinear plasmonic structures as well as graphene-based structures \cite{dai2015low, daneshfar2017switching,sharif2016experimental,naseri2018optical}. To confirm their existence in a PTFBG in the presence of inhomogeneously decreasing nonlinearity, numerical experiments were conducted at $g= 3.5$. If the rate at which the nonlinearity is taperd is small ($\sigma = 3$), then the ramp-like first stable states turn out for a wide span of input intensities. If the nonlinearity decreases at a higher rate (say $\sigma >3$), the span of ramp-like states reduces considerably as shown in Fig. \ref{fig2}(a). The span decreases further in the positive detuning regime which means that the switching to the next stable state occurs at low critical switching intensities as shown in Fig. \ref{fig2}(b). Operating the PTFBGs in the broken regime at different wavelengths ($\delta$) generates  distinct OB curves. These curves are well separated from each other by a definite value of input power. But there exists a range of input intensities common to different spectral components.  Apart from this, the second stable states (corresponding to different $\delta$ values) are well separated from each other in terms of output intensities. These observations are sufficient to conclude that broken $\mathcal{PT}$-symmetric regime is a good choice to generate OB curves with the high spectral uniformity.

\subsection{OB in the unbroken $\mathcal{PT}$-symmetric regime: Right incidence}
\begin{figure}[ht]
	\centering	\includegraphics[width=0.4\linewidth]{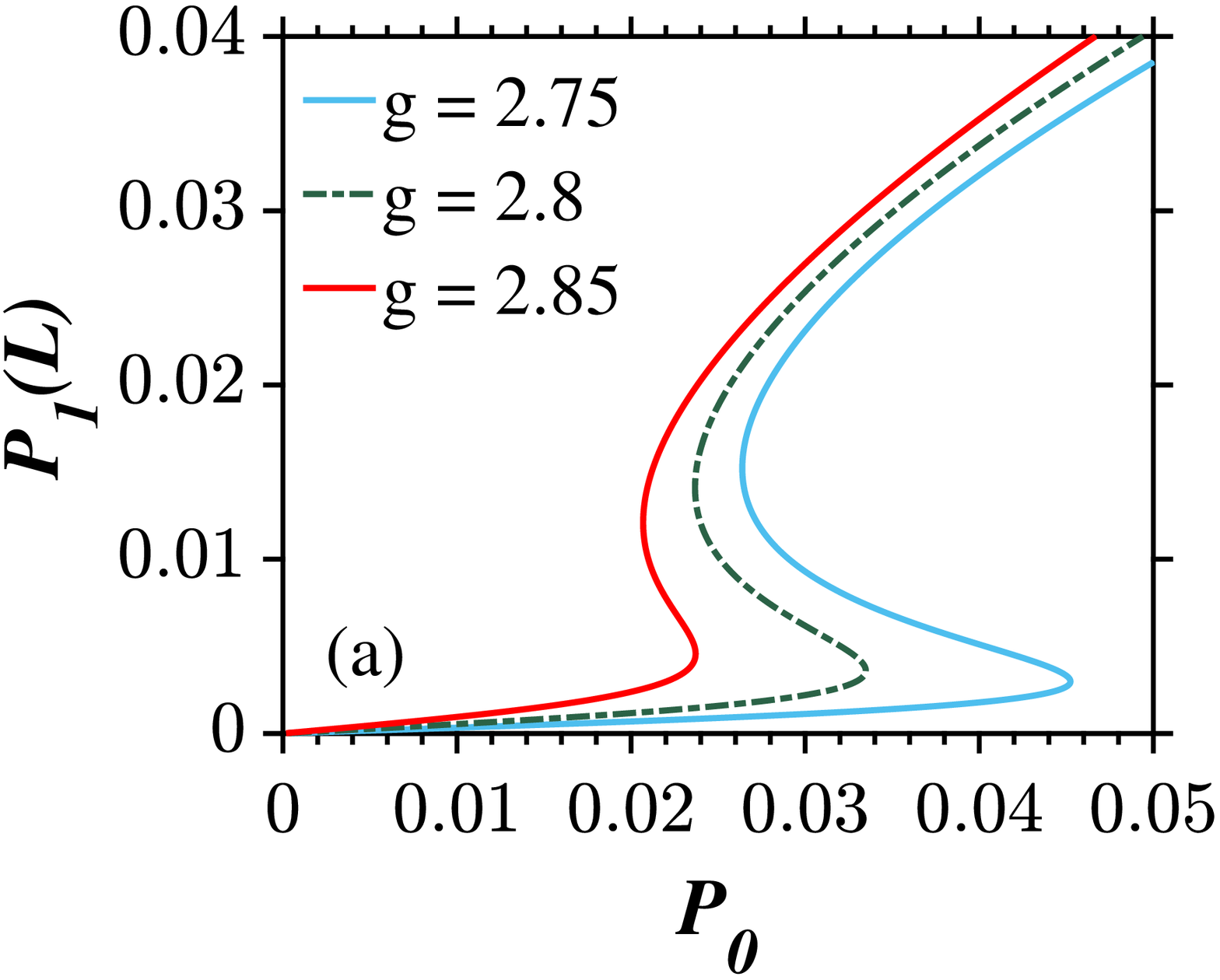}\centering	\includegraphics[width=0.4\linewidth]{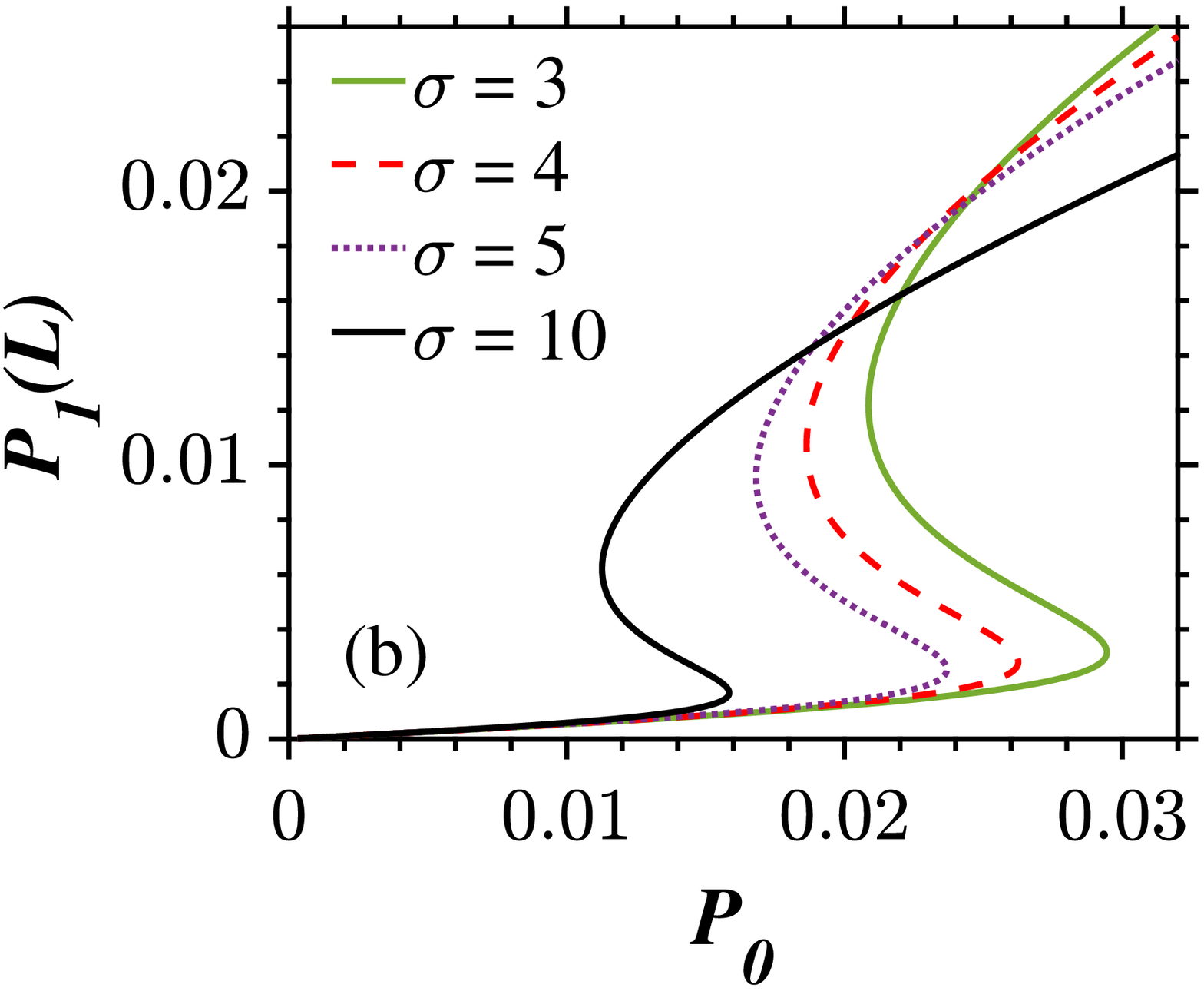}\\\includegraphics[width=0.4\linewidth]{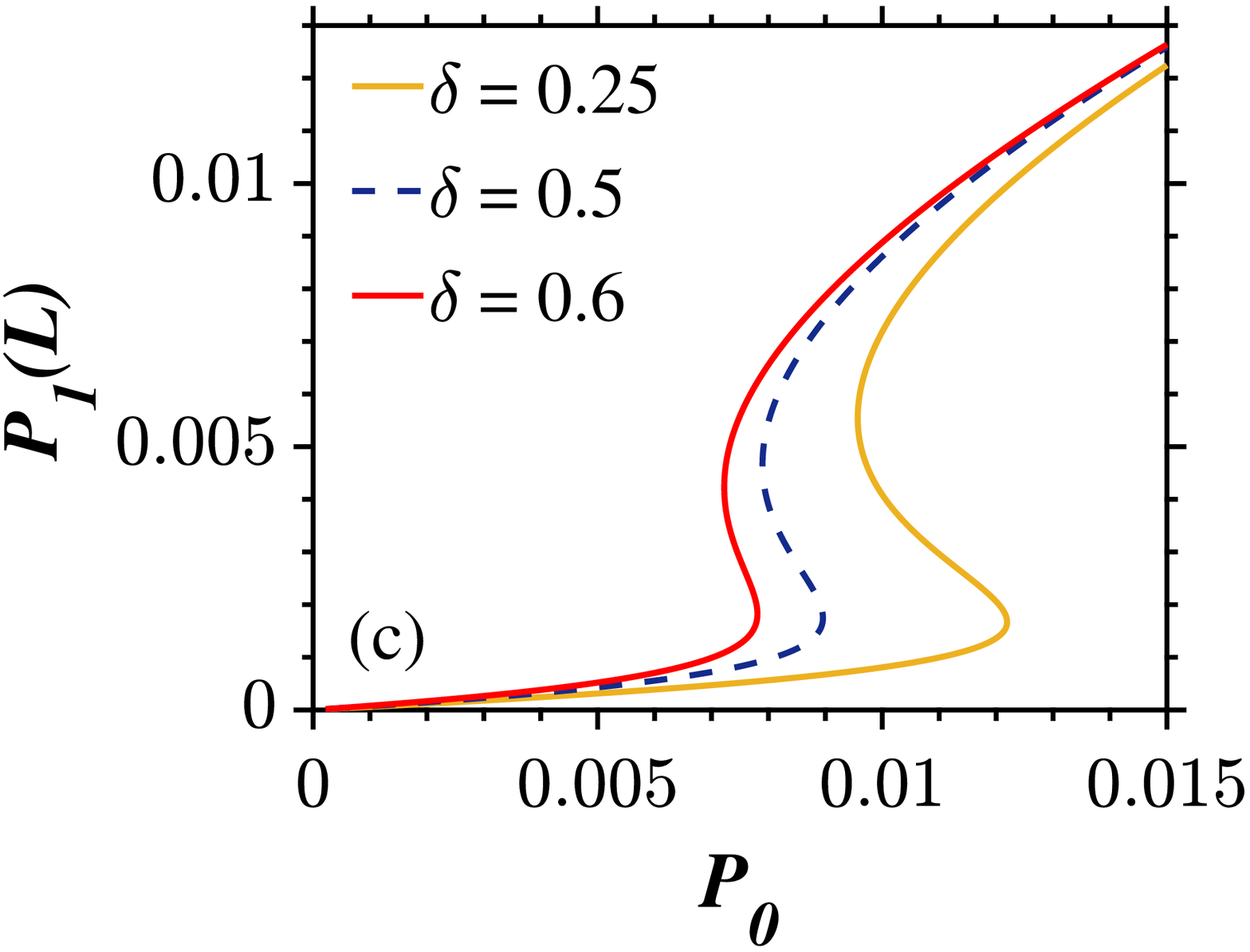}
	\caption{(a)-- (c) Plots portraying the same dynamics with same system parameters as in Fig. \ref{fig1}(a), (c) and (d), respectively except that the light launching direction is right. Also, $\sigma = 10$ in (c).}
	\label{fig3}
\end{figure}
From the fundamental of nonlinear PTFBGs, the term \textit{nonreciprocal switching} refers to the occurrence of distinguishable switching dynamics under the reversal of direction of light incidence under similar operating conditions \cite{raja2019multifaceted,PhysRevA.100.053806,sudhakar2022low,komissarova2019pt}. Changing the light direction also leads to the steering at low critical intensities. What makes the concept of nonreciprocal switching far more competitive than the existing scheme to realize a low-power AOS is that the scheme is straightforward to implement. The underlying physical process responsible for the reduction in the critical switching intensities under reversal of direction of incidence is not fully understood. But the theory of local field enhancement proposed by Kulishov \emph{et al.} provides a closer insight \cite{kulishov2005nonreciprocal}. They suggested that the electric field maxima are located in the gain regions of PTFBG in the presence of right light incidence direction. As a consequence, lasing behavior occurs in the linear regime. It is expected that the same mechanism is responsible for the dramatic reduction in the switching intensities in the nonlinear regime. It can be asked  whether the low-power switching will persist even in the presence of the inhomogeneous nonlinear profile?  Surprisingly, the PTFBG with the linearly decreasing inhomogeneous nonlinearity supports low power OB curves for a wide range of operating conditions as shown in Fig. \ref{fig3}. 

Figure \ref{fig3}(a) delineates the role of gain and loss parameter on the reduction in the critical switching intensities. It should be recalled that $g$ should be chosen in a close proximity to $\kappa$. Only then, the gain and loss parameter will play a beneficial role on the low-power steering dynamics. Under these conditions, an increase in the value of gain and loss  provokes the minimization of $P_{cr}^{\uparrow}$ and $P_{cr}^{\downarrow}$. Further increase in the value of $g$ reduces the chances for the formation of desirable OB states. As an alternate option, we can vary the inhomogeneous nonlinearity parameter ($\sigma$) at fixed values of $g$. From Fig. \ref{fig3}(b), we conclude that the larger the rate at which the nonlinearity is inhomogeneously decreasing, the smaller is the intensity required for the switching. Previously, theoreticians have observed desirable steering dynamics at $\sigma>5$ under different operating conditions \cite{pinto2017bistability,nobrega1998optimum,nobrega2000multistable}. In the absence of $\mathcal{PT}$-symmetry ($g= 0$), decreasing the inhomogeneous profile at a larger rate leads to a loss of desirable OB states at $L =2$ and $\kappa = 3$ (not shown here). But the presence of $\mathcal{PT}$-symmetry ensures that the OB states appear in the transfer characteristics even at larger values of $\sigma$ for the given set of (theoretical) system parameters. The detuning paramter has the potential to reduce the critical switching intensities provided that its sign matches with the type of nonlinearity \cite{raja2019multifaceted,PhysRevA.100.053806,sudhakar2022low}. Along these lines, the PTFBG is operated in the positive detuning regime in the numerical experiments portrayed in Fig. \ref{fig3}(c).  As the magnitude of $\delta$ increases, a steep decline in the critical switching intensities is visible from the plots. For instance, the curves plotted at $\delta = 0.5$ and $0.6$ feature $P_{cr}^{\uparrow\downarrow}<0.01$ at $\sigma = 10$ which must be regarded as the lowest critical intensities ever recorded in the context of PTFBGs. 
\subsection{OB in the broken $\mathcal{PT}$-symmetric  regime: Right incidence}
\begin{figure}[t]
	\centering	\includegraphics[width=0.4\linewidth]{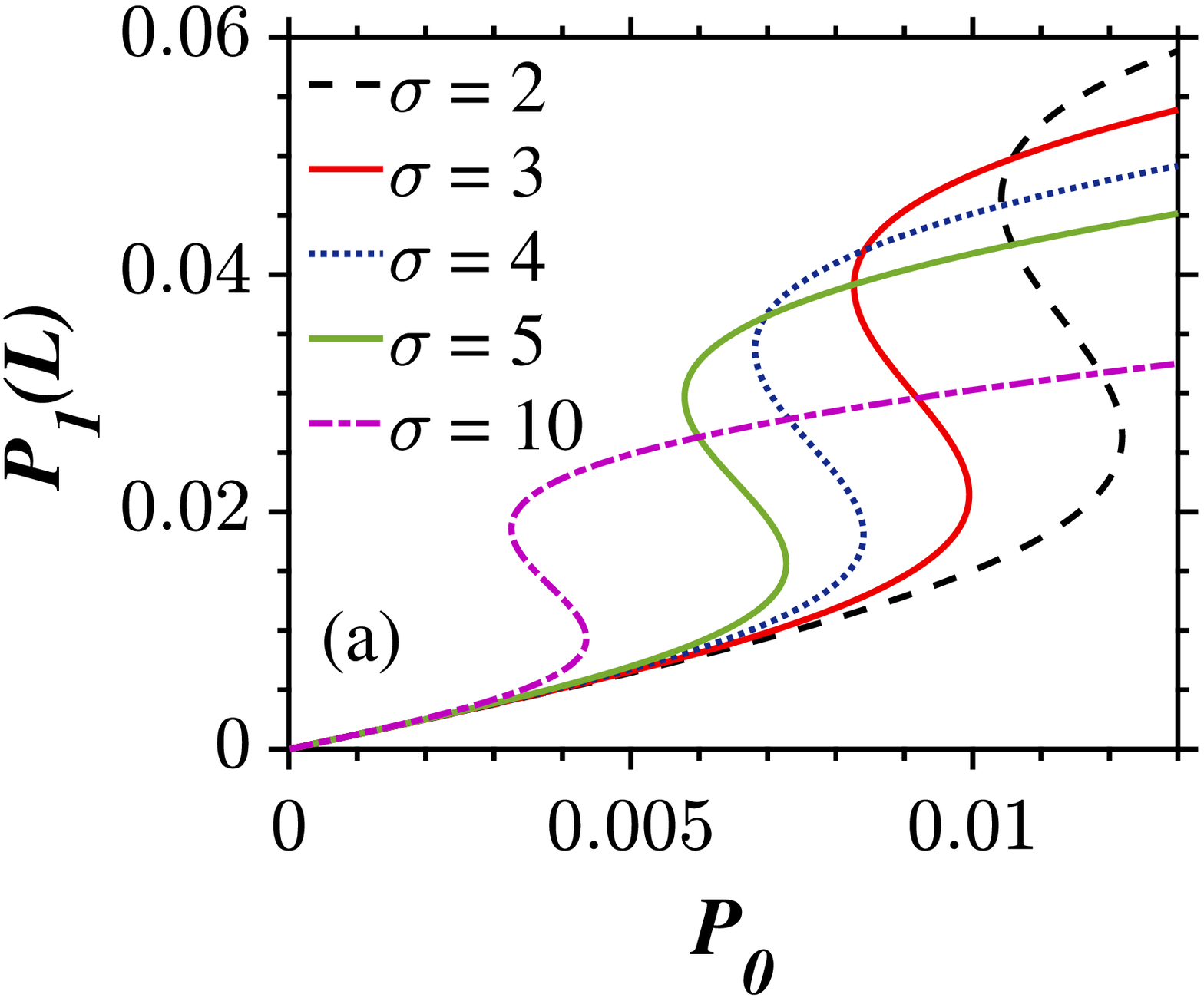}\centering	\includegraphics[width=0.4\linewidth]{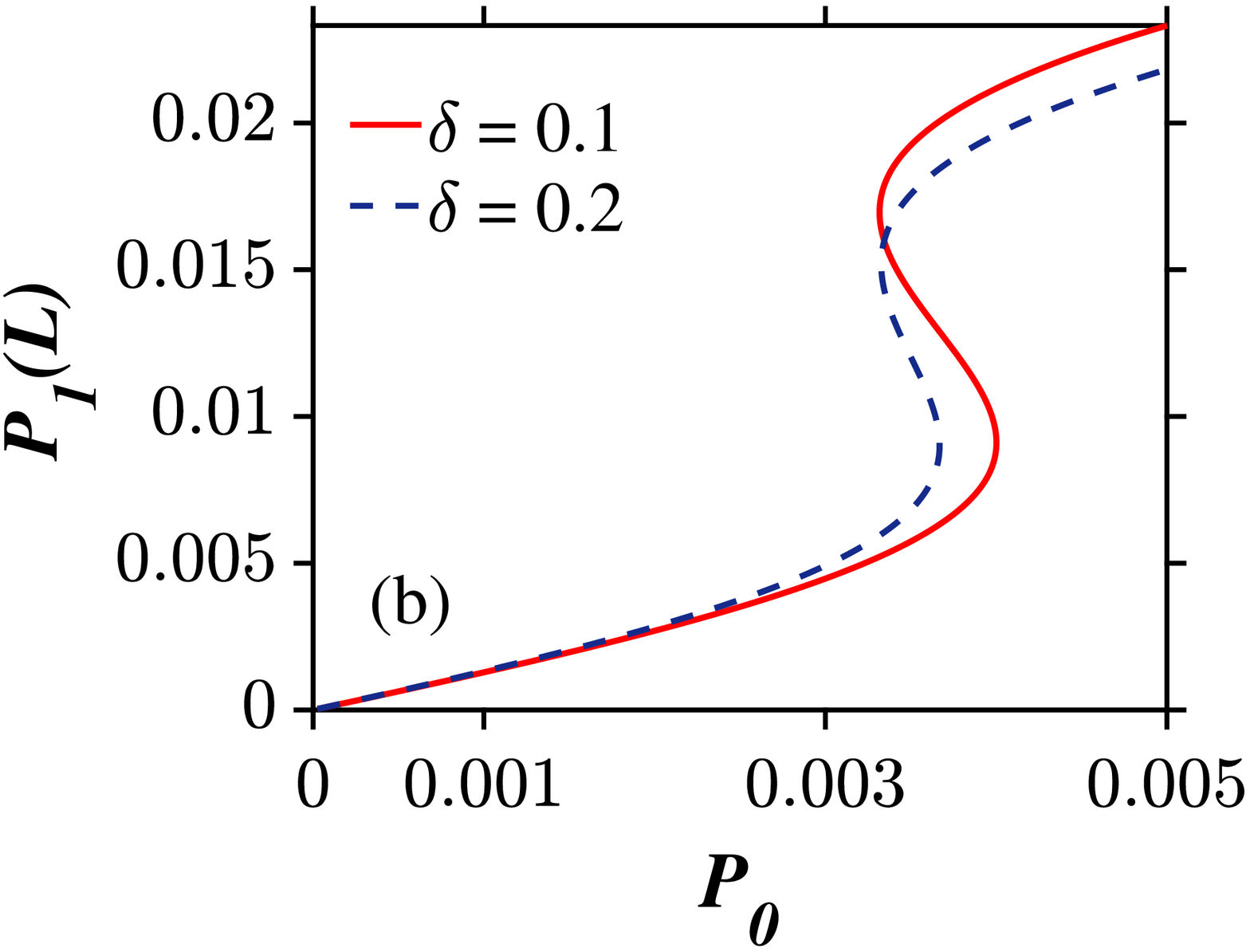}
	\caption{Ultra-low power steering in a broken PTFBG ($g=3.5$) with linearly decreasing inhomogeneous nonlinearity. The direction of light incidence is right. (a) Role of gain and loss at $\delta= 0$ and (b) positive detuning parameter ($\delta$) at $\sigma = 10$ on the switching dynamics.}
	\label{fig4}
\end{figure}
\begin{figure}
	\centering	\includegraphics[width=0.4\linewidth]{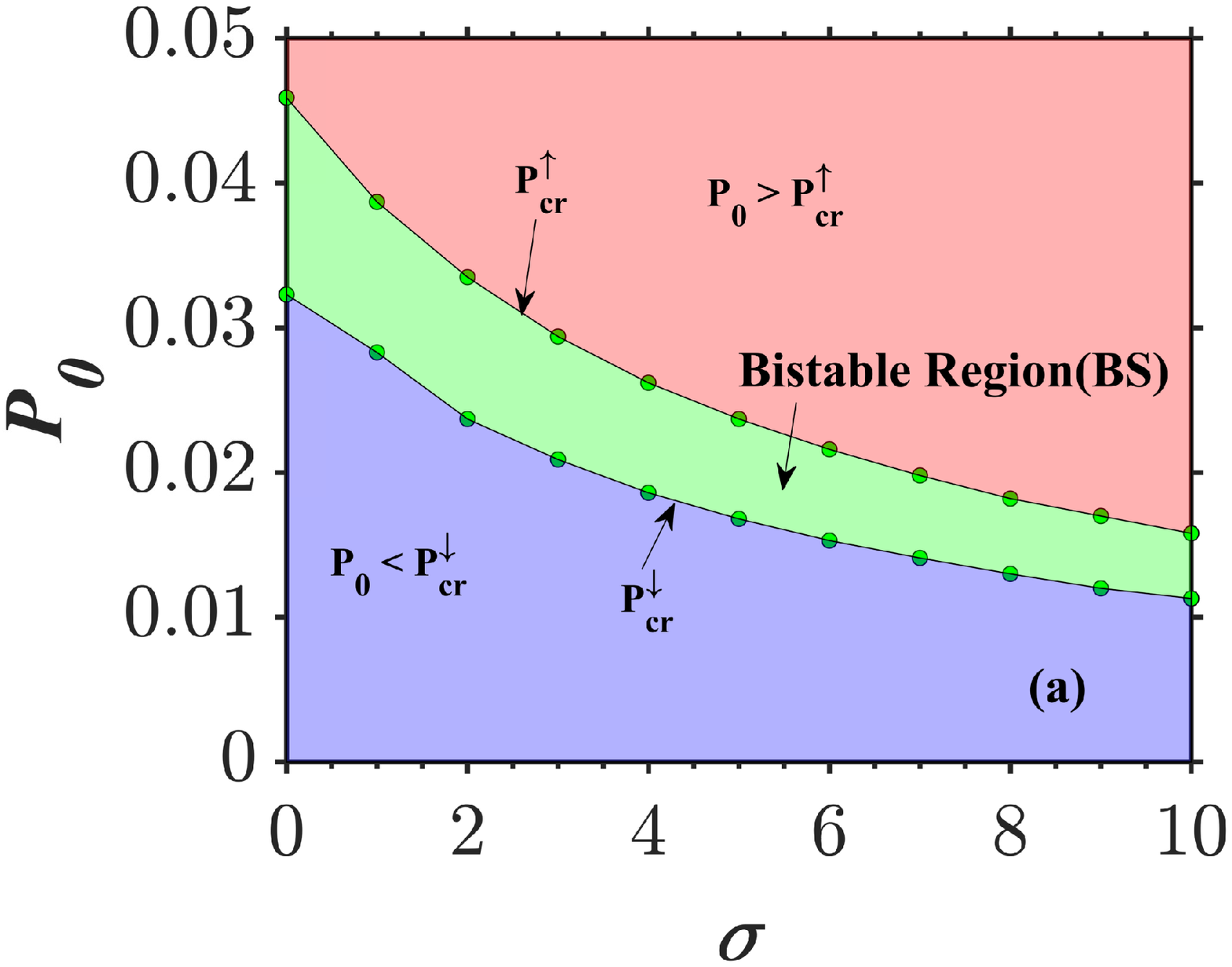}\centering	\includegraphics[width=0.4\linewidth]{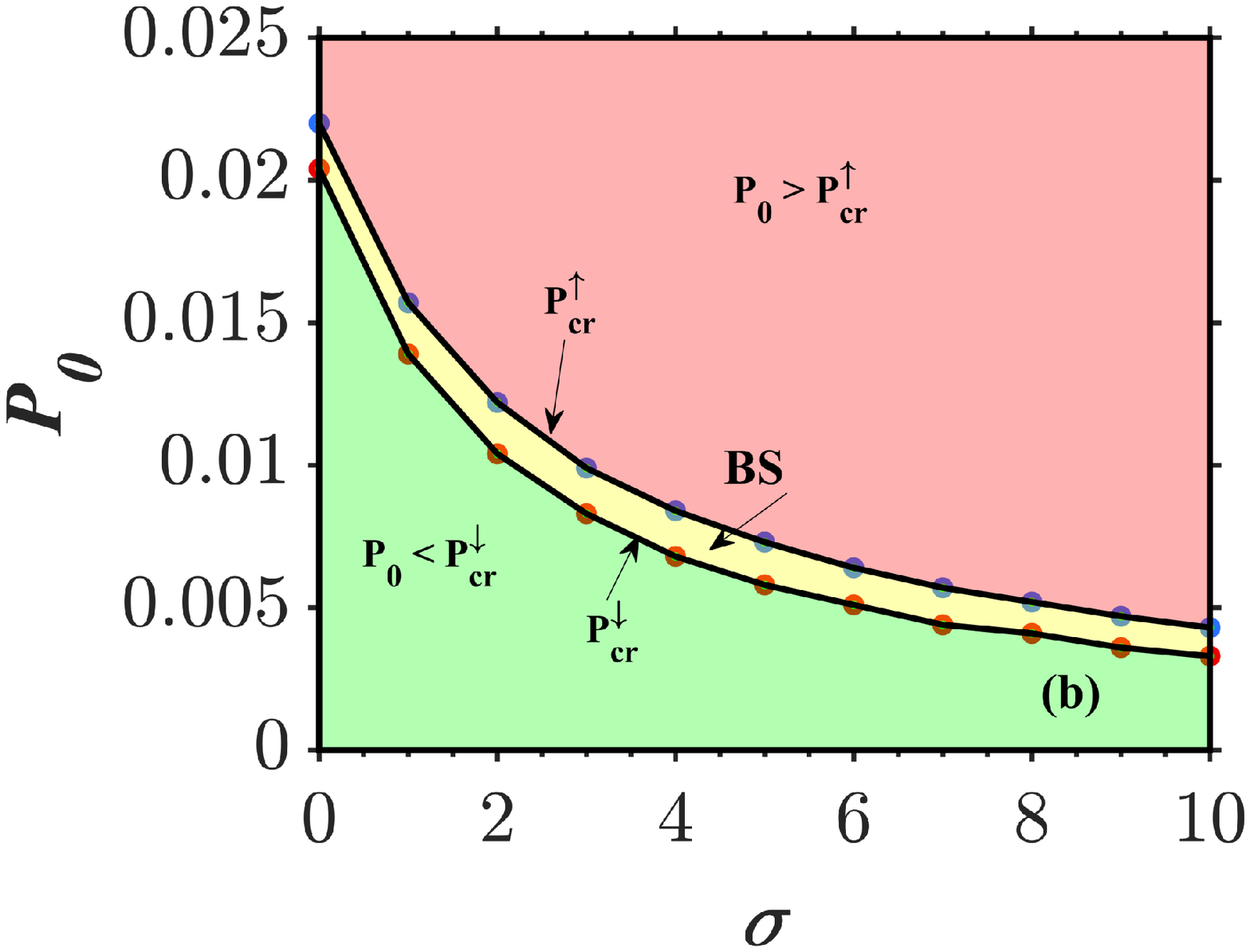}
	\caption{Role of inhomogeneous nonlinearity parameter ($\sigma$) in the ultra-low power switching dynamics exhibited by an (a) unbroken ($g = 2.8$) and (b) broken PTFBG ($g= 3.5$). The direction of light incidence is right.}
	\label{cvfig}
\end{figure}
The proposed system supports the phenomenon of low-power steering in the broken regime too as confirmed from the plots shown in Fig. \ref{fig4}. What differentiates the present system from previously investigated PTFBG systems \cite{raja2019multifaceted,PhysRevA.100.053806,sudhakar2022low} is that the difference between the switching intensities pertaining to the unbroken and the broken regimes under similar operating conditions. It is found that the intensity required to switch in the broken regime is always higher than that of the intensity in the unbroken regime in the existing systems. This inference holds good for a wide range of system parameters, namely chirping \cite{PhysRevA.100.053806}, detuning \cite{raja2019multifaceted, PhysRevA.100.053806}, modulation of Kerr nonlinearity \cite{sudhakar2022low} and higher-order nonlinearities \cite{raja2019multifaceted}.  For instance, in Ref. \cite{sudhakar2022low} an effort to narrow down the switching intensities in the broken regime via the frequency detuning leads to the loss of bistability itself. But the proposed system behaves in a different fashion. It is possible to achieve ultra-low power switching in the broken regime under the reversal of direction of light incidence, thanks to the concept of inhomogeneous decreasing nonlinearity. The critical switching intensities of unbroken and broken PTFBGs in the right light incidence condition are compared in Fig. \ref{cvfig}. From the figures, it is very clear that the broken PTFBG features lower intensities than unbroken PTFBG provided that other system parameters are unaltered. The broken regime which was once considered as the instability regime \cite{komissarova2019pt} is now proven to be the potential regime to realize low-power AOS. Theoretically, it is possible to reduce the critical intensities to a value less than 0.002 (plots not shown here) when the broken PTFBG is operated at the synchronous wavelength with a high value of inhomogeneous nonlinearity ($\sigma >15$) .

\subsection{Peculiar OB curves in the broken regime: Increasing inhomogeneous nonlinearity}
\begin{figure}
	\centering	\includegraphics[width=0.5\linewidth]{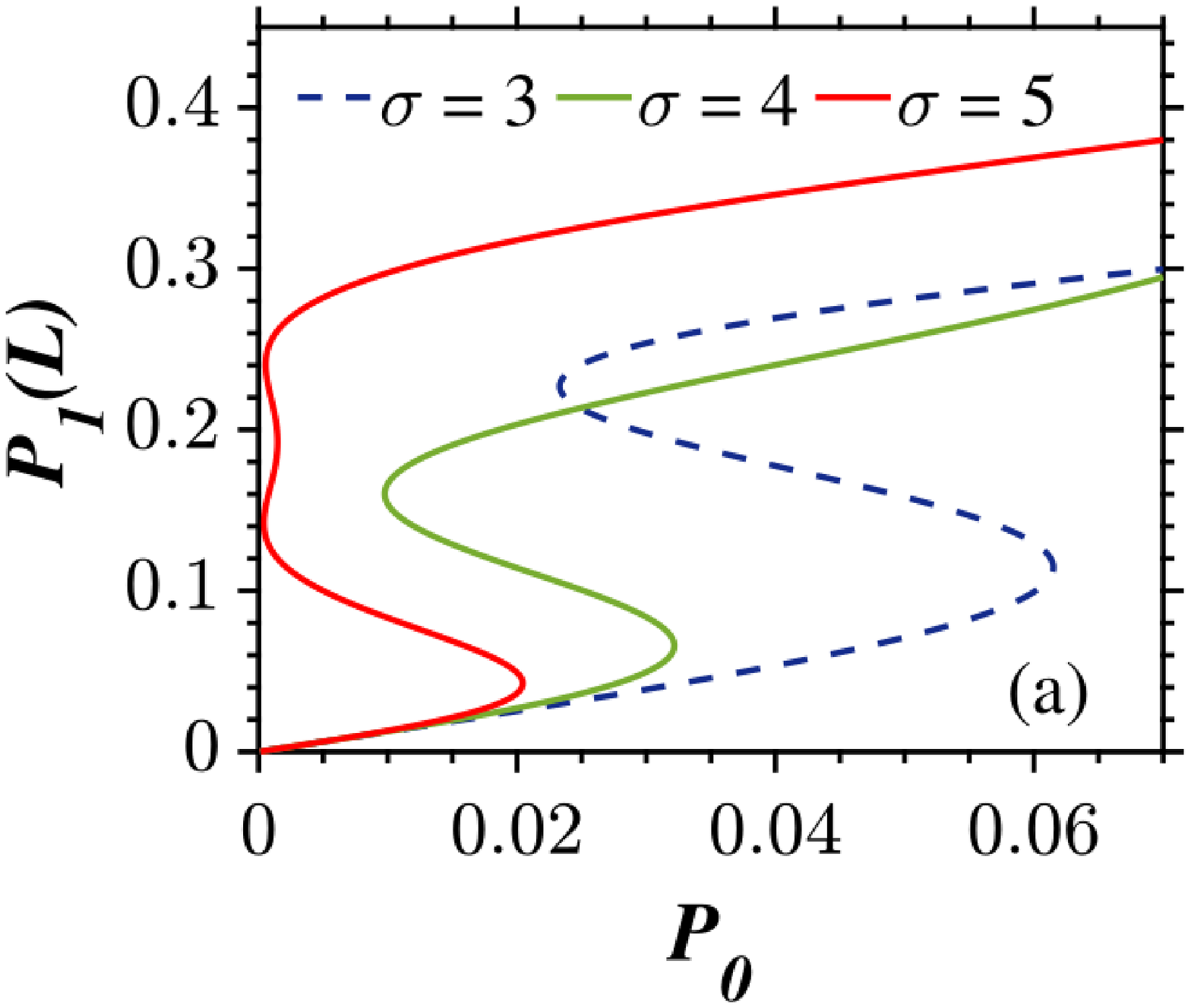}\centering	\includegraphics[width=0.5\linewidth]{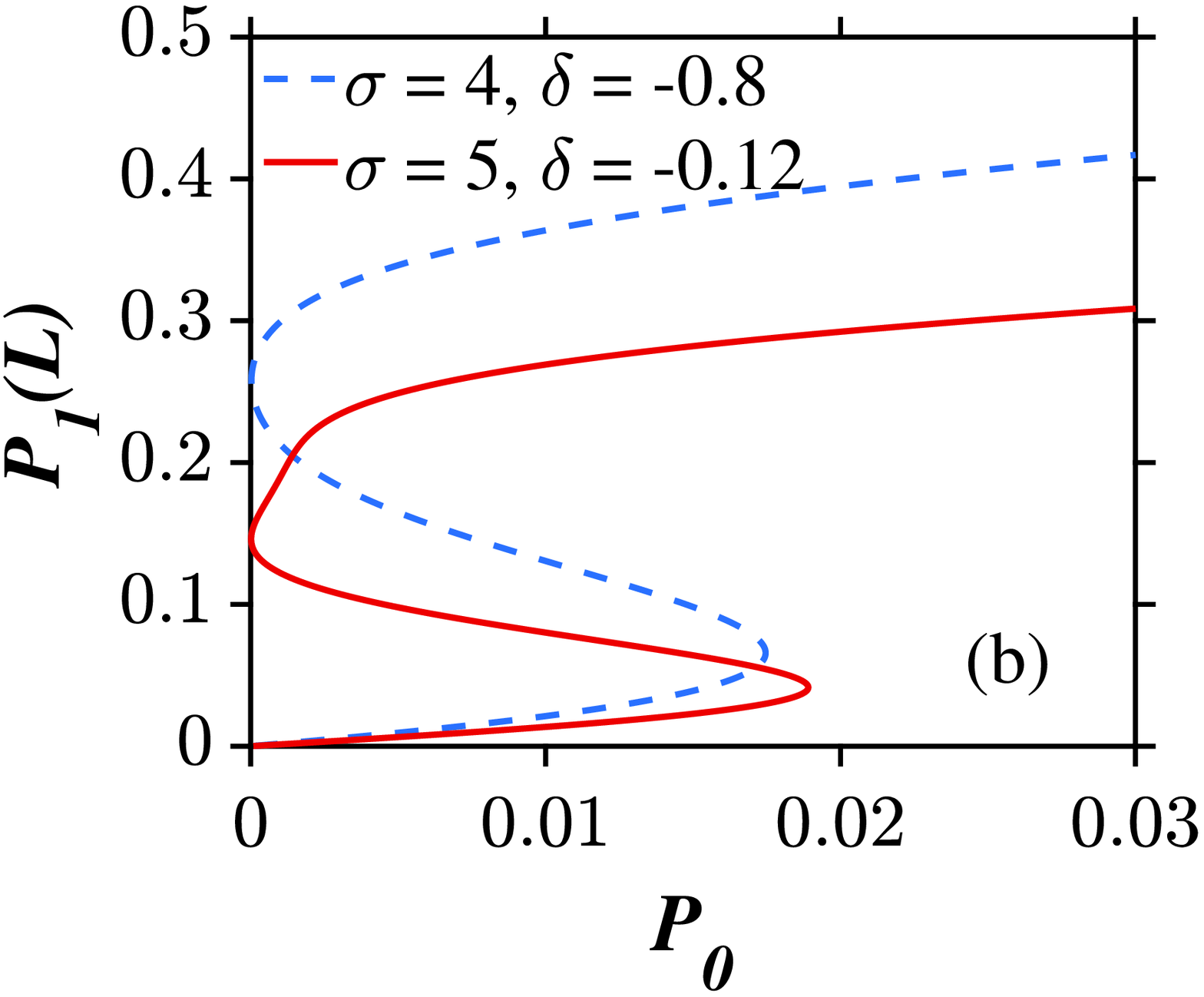}
	\caption{OB in a broken PTFBG ($g =3.5$) with linearly increasing inhomogeneous nonlinearity under right light incidence direction at $\delta = 0$. (a) System is operated at the synchronous wavelength ($\delta = 0$). (b) OB curves featuring near-zero critical switch-down intensities ($P_{cr}^{\downarrow}$).}
	\label{fig5}
\end{figure}
Under the reversal of direction of incidence, the broken PTFBG with the linearly increasing inhomogeneous nonlinearity generates OB with zero switch-down intensities ($P_{cr}^{\downarrow}$) at higher values of $\sigma$ as shown in Fig. \ref{fig5}(a). This means that the system can return back to its original states only at zero intensity. The second stable state gets flattened and remains in the same shape for a wide span of input intensities. The system does not favor the formation of OM even at higher intensities. This means that the operation of the device is bistable for a wide range of intensities. This kind of OB curves with zero switch-down intensities were known to exist in plasmonic devices with nanoparticles \cite{daneshfar2017switching}, $\mathcal{PT}$-symmetric photonic crystals \cite{xu2022optical}, Bose-Einstein condensate optomechanical systems \cite{yasir2015tunable}, and anti-directional couplers \cite{govindarajan2019nonlinear}. Even though the switch-down action takes place at zero-intensities, the corresponding switch-up intensities are higher in the reported works \cite{daneshfar2017switching,xu2022optical,yasir2015tunable,govindarajan2019nonlinear}. But in the present investigations, the critical switch-up intensity is also low besides the occurrence of switch-down action at zero-intensities.   This is a remarkable finding in the context of PTFBGs.   OB states with a low-switch, zero switch-down intensities and flat second stable states can be used as two level memories or binary flip-flops as they can stay in the second states for a larger input intensity values. The critical intensities can function as read and write bias values in the context of optical memories.  For lower values of $\sigma$, low-power OB curves with ramp-like stable states are obtained.    Fig. \ref{fig5}(b) confirms that OB states with zero critical switch-down intensities can be obtained at lower values of inhomogeneosuly decreasing nonlinearity parameter ($\sigma = 4$) under a suitable manipulation of the detuning parameter. 

\section{Summary and conclusions}
The article starts off with the modeling and investigation on the role of $\mathcal{PT}$-symmetry on the switching dynamics of inhomogeneous PTFBGs and it is found that closer the value of gain and loss parameter, smaller is the intensity required to switch. The inhomogeneously increasing nonlinearity profile is found to generate OB curves at higher intensities. On the contrary, decreasing nonlinear profile is found to be an optimum choice for generating OB curves at low intensities. Investigations in the unbroken and broken regimes with conventional light launching conditions reveal that the critical switching intensities can be brought down by controlling the rate of inhomogeneously decreasing nonlinearity or operating in the positive detuning regime. The formation of ramp-like bistable states persists in the broken regime even though the nonlinear profile is inhomogeneous. Significant discoveries were made in the unbroken as well as broken right regimes under the reversal of direction of incidence. It is found that the critical switching intensities can be trimmed to a value $<$ 0.01 provided that the value of inhomogeneously decreasing nonlinearity is very high. The broken regime that usually generates OB curves at higher intensities than the unbroken regime works in an unconventional manner. In other words, the critical switching intensities are lower in the broken regime than the switching intensities in the unbroken regime. In fact, the operation of the proposed system in the broken regime features the lowest critical switching intensities in the context of PTFBGs. Remarkably, the low intensity bistable curves show ramp-like stable states before the onset of switching. In the presence of right light incidence direction, operating the PTFBG with the linearly increasing inhomogeneous nonlinearity in the broken regime generates OB curves with \textit{zero critical switching} intensities. These numerical results deserve a practical realization through experiments. As theoretical physicists, we believe  that inhomogeneous doping of the core material will help in the practical realization of FBGs with the inhomogeneous nonlinear profile. This may open a new avenue to control light with light in the next generation all-optical signal processing devices and AOS.

\begin{backmatter}
\bmsection{Acknowledgements}
SVR is supported by the  Department of Science and Technology (DST)-Science and Engineering Research Board (SERB), Government of India, through a National Postdoctoral Fellowship (Grant
No. PDF/2021/004579). AG is supported by the University Grants Commission (UGC), Government of India,
through a Dr. D. S. Kothari Postdoctoral Fellowship (Grant
No. F.4-2/2006 (BSR)/PH/19-20/0025). ML is supported by a DST-SERB National Science Chair.

\

\bmsection{Disclosures}
\noindent The authors declare no conflicts of interest.

\bmsection{Data Availability Statement}
 No data were generated or analyzed in the presented research.

\end{backmatter}

\end{document}